Gauge invariance and the vacuum state

by

Dan Solomon
Rauland-Borg Corporation
3450 W. Oakton
Skokie, IL  60076

Please send all correspondence to:

Dan Solomon
1604 Brummel
Evanston, IL  60202

Phone: 847-679-0900 ext. 5337
Email: dsolomon@northshore.net

May 4, 1999

## Abstract

Quantum field theory is assumed to be gauge invariant.  It is shown that for a Dirac field the assumption of gauge invariance impacts on the way the vacuum state is defined.  It is shown that the conventional definition of the vacuum state must be modified to take into account the requirements of gauge invariance.

<u>I. Introduction</u>

Quantum field theory is assumed to be gauge invariant (see Schwinger [1]).  It will be shown in this article that in Dirac field theory the requirement of gauge invariance puts some requirements on the way the vacuum state vector $|0\rangle$ is defined.  It will be shown that the conventional definition of the vacuum state must be modified to take into account the requirements of gauge invariance.  Throughout this discussion natural units are used so that $c = \hbar = 1$.

In the Schrodinger representation of Dirac field theory the time evolution of the state vector $|\Omega(t)\rangle$ and its dual $\langle\Omega(t)|$ is given by

$$i\frac{\partial|\Omega(t)\rangle}{\partial t} = \hat{H}(t)|\Omega(t)\rangle; \quad -i\frac{\partial\langle\Omega(t)|}{\partial t} = \langle\Omega(t)|\hat{H}(t) \qquad (1)$$

where $\hat{H}(t)$ is the Hamiltonian operator and is given by

$$\hat{H}(t) = \hat{H}_o - \int\hat{\vec{J}}(\vec{x})\cdot\vec{A}(\vec{x},t)d\vec{x} + \int\hat{\rho}(\vec{x})A_o(\vec{x},t)d\vec{x} \qquad (2)$$

(See, for example, Rayski[2] or sections 17 and 19 of Pauli[3].)  In the above expression $\left(A_o(\vec{x},t), \vec{A}(\vec{x},t)\right)$ is the electric potential.  In this article the electric potentials are assumed to be unquantized, real valued functions.  $\hat{\vec{J}}(\vec{x})$ and $\hat{\rho}(\vec{x})$ are the current and charge operators, respectively and $\hat{H}_o$ is the free field Hamiltonian, which is the Hamiltonian operator when the interactions are turned off (i.e. the electric potential is zero).  Throughout this discussion it is assumed that $|\Omega(t)\rangle$ is normalized, i.e.,

$$\langle\Omega(t)|\Omega(t)\rangle = 1.$$

The free field energy is defined by

$$\xi_o(t) = \left\langle \Omega(t) \middle| \hat{H}_o \middle| \Omega(t) \right\rangle \tag{3}$$

Now given the quantum state $\left| \Omega(t_a) \right\rangle$, defined at t=$t_a$, the application of Eq. (1) will evolve the quantum state forward in time to produce the quantum state $\left| \Omega(t_b) \right\rangle$ at t=$t_b$. In general $\left| \Omega(t_b) \right\rangle \neq \left| \Omega(t_a) \right\rangle$ and, if the electric potential is not zero, then, in general, $\xi_o(t_a) \neq \xi_o(t_b)$. The question that will be addressed in this article is whether or not there is a lower bound to $\xi_o(t)$. It will be shown that if quantum theory is gauge invariant then the answer to this question is no. There is no lower bound to $\xi_o(t)$. Or to put the problem more precisely, it will be shown that, given the quantum state $\left| \Omega(t_a) \right\rangle$, defined at t=$t_a$, it is possible to find an electric potential $\left( A_o(\vec{x},t), \vec{A}(\vec{x},t) \right)$ that, when used in Eq (1), evolves the quantum state into $\left| \Omega(t_b) \right\rangle$ at t=$t_b$ such that $\xi_o(t_b)$ is less then $\xi_o(t_a)$ by an arbitrarily large amount.

## II. Gauge invariance

The electromagnetic field is given in terms of the electric potential according to

$$\vec{E} = -\left( \frac{\partial \vec{A}}{\partial t} + \vec{\nabla} A_o \right); \quad \vec{B} = \vec{\nabla} \times \vec{A} \tag{4}$$

A change in the gauge is a change in the electric potential that produces no change in the electromagnetic field. Such a change is given by

$$\vec{A} \rightarrow \vec{A}' = \vec{A} - \vec{\nabla}\chi; \quad A_o \rightarrow A_o' = A_o + \frac{\partial \chi}{\partial t} \tag{5}$$

where $\chi(\vec{x}, t)$ is an arbitrary real valued function.

Now, in general, when a change in the gauge is introduced into Eq. (1) this will produce a change in $\left|\Omega(t)\right\rangle$. However, an important requirement of a physical theory is that it be gauge invariant which means that the change in the gauge produces no change in the physical observables. These include the current and charge expectation values which are defined by

$$\vec{J}_e(\vec{x}, t) = \left\langle \Omega(t) \left| \hat{\vec{J}}(\vec{x}) \right| \Omega(t) \right\rangle \tag{6}$$

and

$$\rho_e(\vec{x}, t) = \left\langle \Omega(t) \left| \hat{\rho}(\vec{x}) \right| \Omega(t) \right\rangle \tag{7}$$

The requirement for gauge invariance can be expressed as follows. Let there exist two state vectors $\left|\Omega_1(t)\right\rangle$ and $\left|\Omega_2(t)\right\rangle$ which obey Eq. (1). The electric potential that acts on $\left|\Omega_1(t)\right\rangle$ is $\left(A_o^{(1)}(\vec{x}, t), \vec{A}^{(1)}(\vec{x}, t)\right)$ and the electric potential that acts on $\left|\Omega_2(t)\right\rangle$ is $\left(A_o^{(2)}(\vec{x}, t), \vec{A}^{(2)}(\vec{x}, t)\right)$. Assume the electric potentials are related by a gauge transformation, that is

$$\left(A_o^{(2)}(\vec{x}, t), \vec{A}^{(2)}(\vec{x}, t)\right) = \left(A_o^{(1)}(\vec{x}, t) + \frac{\partial \chi(\vec{x}, t)}{\partial t}, \vec{A}^{(1)}(\vec{x}, t) - \vec{\nabla}\chi(\vec{x}, t)\right) \tag{8}$$

Assume that at $t = t_a$ we have that

$$\left|\Omega_1(t_a)\right\rangle = \left|\Omega_2(t_a)\right\rangle \tag{9}$$

and

$$\chi(\vec{x}, t_a) = 0 \text{ and } \frac{\partial \chi(\vec{x}, t_a)}{\partial t} = 0 \tag{10}$$

Other then this constraint $\chi(\vec{x}, t)$ is an arbitrary real valued function.

We have defined two states, which are equal at time $t=t_a$ but evolve forward in time under the action of electric potentials which are related by a gauge transformation. Then, even though in general $\left| \Omega_1(t) \right\rangle \neq \left| \Omega_2(t) \right\rangle$ for $t>t_a$ , gauge invariance requires that the physically observable quantities for each quantum state be equal, that is,

$$\vec{J}_e^{(2)}(\vec{x}, t) = \vec{J}_e^{(1)}(\vec{x}, t) \text{ for } t \geq t_a \tag{11}$$

and

$$\rho_e^{(2)}(\vec{x}, t) = \rho_e^{(1)}(\vec{x}, t) \text{ for } t \geq t_a \tag{12}$$

where $\vec{J}_e^{(r)}(\vec{x}, t)$ and $\rho_e^{(r)}(\vec{x}, t)$ are the current and charge expectation value, respectively, for the quantum state $\left| \Omega_r(t) \right\rangle$ where r=1,2.

Next we will derive an expression for the time derivative of $\xi_o(t)$ which will be used later in the discussion.  Use Eq. (2) in (3) to obtain

$$\xi_o(t) = \left\langle \Omega \left| \left\{ \hat{H} - \left( -\int \hat{\vec{J}}(\vec{x}) \cdot \vec{A}(\vec{x}, t) d\vec{x} + \int \hat{\rho}(\vec{x}) A_o(\vec{x}, t) d\vec{x} \right) \right\} \right| \Omega \right\rangle \tag{13}$$

Use (6) and (7) in the above to obtain

$$\xi_o(t) = \left\langle \Omega \left| \hat{H}(t) \right| \Omega \right\rangle - \left\{ \left( -\int \vec{J}_e(\vec{x}, t) \cdot \vec{A}(\vec{x}, t) d\vec{x} + \int \rho_e(\vec{x}, t) A_o(\vec{x}, t) d\vec{x} \right) \right\} \tag{14}$$

Take the derivative of the above expression with respect to time to obtain

$$\frac{d\xi_o(t)}{dt} = \frac{d}{dt} \left\langle \Omega \left| \hat{H}(t) \right| \Omega \right\rangle - \left\{ \left( -\int \frac{\partial \vec{J}_e(\vec{x}, t)}{\partial t} \cdot \vec{A}(\vec{x}, t) d\vec{x} + \int \frac{\partial \rho_e(\vec{x}, t)}{\partial t} A_o(\vec{x}, t) d\vec{x} \right) \right\}$$
$$- \left( -\int \vec{J}_e(\vec{x}, t) \cdot \frac{\partial \vec{A}(\vec{x}, t)}{\partial t} d\vec{x} + \int \rho_e(\vec{x}, t) \frac{\partial A_o(\vec{x}, t)}{\partial t} d\vec{x} \right) \tag{15}$$

Now the time derivative of $\langle\Omega|\hat{H}(t)|\Omega\rangle$ is given by

$$\frac{d}{dt}\langle\Omega|\hat{H}(t)|\Omega\rangle = \left(\frac{\partial\langle\Omega|}{\partial t}\right)\hat{H}(t)|\Omega\rangle + \langle\Omega|\frac{\partial\hat{H}(t)}{\partial t}|\Omega\rangle + \langle\Omega|\hat{H}(t)\left(\frac{\partial|\Omega\rangle}{\partial t}\right) \qquad (16)$$

Use Eqs (1) and (2) in the above to yield

$$\frac{d}{dt}\langle\Omega|\hat{H}(t)|\Omega\rangle = \langle\Omega|\frac{\partial\hat{H}(t)}{\partial t}|\Omega\rangle = \langle\Omega|\left\{\left(-\int\hat{\vec{J}}(\vec{x})\cdot\frac{\partial\vec{A}(\vec{x},t)}{\partial t}d\vec{x} + \int\hat{\rho}(\vec{x})\frac{\partial A_o(\vec{x},t)}{\partial t}d\vec{x}\right)\right\}|\Omega\rangle$$

$$(17)$$

Use (6) and (7) in the above to yield

$$\frac{d}{dt}\langle\Omega|\hat{H}(t)|\Omega\rangle = -\int\vec{J}_e(\vec{x},t)\cdot\frac{\partial\vec{A}(\vec{x},t)}{\partial t}d\vec{x} + \int\rho_e(\vec{x},t)\frac{\partial A_o(\vec{x},t)}{\partial t}d\vec{x} \qquad (18)$$

Use this in Eq. (15) to obtain

$$\frac{d\xi_o}{dt} = \int\frac{\partial\vec{J}_e(\vec{x},t)}{\partial t}\cdot\vec{A}(\vec{x},t)d\vec{x} - \int\frac{\partial\rho_e(\vec{x},t)}{\partial t}A_o(\vec{x},t)d\vec{x} \qquad (19)$$

### III Negative free field energy

In this section it will be shown that if quantum theory is gauge invariant then there is no lower bound to the free field energy.

Assume the existence of a quantum state $|\Omega_1(t)\rangle$. The electric potential, $\left(A_o^{(1)}(\vec{x},t),\vec{A}^{(1)}(\vec{x},t)\right)$, acting on $|\Omega_1(t)\rangle$ is zero, i.e.,

$$\left(A_o^{(1)}(\vec{x},t),\vec{A}^{(1)}(\vec{x},t)\right) = 0 \qquad (20)$$

Assume that at time $t=t_b$ the charge expectation value for $|\Omega_1(t)\rangle$ satisfies the condition

$$\frac{\partial\rho_e^{(1)}(\vec{x},t)}{\partial t} \neq 0\bigg|_{t=t_b} \qquad (21)$$

Now let us create another quantum state $\left|\Omega_2(t)\right\rangle$ which is derived from $\left|\Omega_1(t)\right\rangle$ in the following manner. At $t=t_a$, where $t_a<t_b$, we have that

$$\left|\Omega_2(t_a)\right\rangle = \left|\Omega_1(t_a)\right\rangle \tag{22}$$

This may be considered an initial condition on $\left|\Omega_2(t)\right\rangle$. Then $\left|\Omega_2(t)\right\rangle$ evolves forward in time according Eq. (1) in the presence of the electric potential

$$\left(A_o^{(2)}(\vec{x},t), \vec{A}^{(2)}(\vec{x},t)\right) = \left(\frac{\partial\chi(\vec{x},t)}{\partial t}, \left(-\vec{\nabla}\chi(\vec{x},t)\right)\right) \tag{23}$$

where $\chi(\vec{x},t)$ is a real valued function subject to the following condition

$$\chi(\vec{x},t_a)=0 \text{ and } \frac{\partial\chi(\vec{x},t_a)}{\partial t}=0 \tag{24}$$

Other then these constraints $\chi(\vec{x},t)$ is arbitrary.

Now how do we know that a quantum state can be found where Eq. (21) is true? If our theory is a correct model of the real world, that is real electrons, then there must be quantum states where Eq. (21) holds because there are numerous examples in the real world where the time derivative of the charge density is not zero.

Given the above, we will prove the following. If quantum theory is gauge invariant then it is possible to find a $\chi(\vec{x},t)$ such that the free field energy $\xi_o^{(2)}(t)$ of the quantum state $\left|\Omega_2(t)\right\rangle$ at time $t=t_b$ is a negative number with an arbitrarily large magnitude.

First, note that, from Eq. (22) we have that

$$\xi_o^{(2)}(t_a) = \xi_o^{(1)}(t_a) \tag{25}$$

Also, referring to Eq. (19), if the electric potential is zero, then $\xi_o(t)$ is constant in time.

Therefore for the quantum state $\left| \Omega_1(t) \right\rangle$,

$$\xi_o^{(1)}(t_b) = \xi_o^{(1)}(t_a) \tag{26}$$

Also, note that $\left( A_o^{(1)}(\vec{x},t), \vec{A}^{(1)}(\vec{x},t) \right)$ and $\left( A_o^{(2)}(\vec{x},t), \vec{A}^{(2)}(\vec{x},t) \right)$ are related by

a gauge transformation (compare Eqs. (20) and (23)). Therefore, if quantum theory is

gauge invariant, then

$$\vec{J}_e^{(1)}(\vec{x},t) = \vec{J}_e^{(2)}(\vec{x},t); \quad \rho_e^{(1)}(\vec{x},t) = \rho_e^{(2)}(\vec{x},t) \tag{27}$$

Next use Eq. (23) in (19) to obtain for the free field energy of $\left| \Omega_2(t) \right\rangle$

$$\frac{d\xi_o^{(2)}}{dt} = -\int \frac{\partial \vec{J}_e^{(2)}(\vec{x},t)}{\partial t} \cdot \vec{\nabla}\chi(\vec{x},t)d\vec{x} - \int \frac{\partial \rho_e^{(2)}(\vec{x},t)}{\partial t}\frac{\partial \chi(\vec{x},t)}{\partial t}d\vec{x} \tag{28}$$

Assume reasonable boundary conditions at $|\vec{x}| \rightarrow \infty$ and integrate by parts to obtain

$$\frac{d\xi_o^{(2)}}{dt} = \int \chi(\vec{x},t)\frac{\partial \vec{\nabla} \cdot \vec{J}_e^{(2)}(\vec{x},t)}{\partial t}d\vec{x} - \int \frac{\partial \rho_e^{(2)}(\vec{x},t)}{\partial t}\frac{\partial \chi(\vec{x},t)}{\partial t}d\vec{x} \tag{29}$$

This can be written as

$$\frac{d\xi_o^{(2)}}{dt} = \int \chi(\vec{x},t)\frac{\partial \vec{\nabla} \cdot \vec{J}_e^{(2)}(\vec{x},t)}{\partial t}d\vec{x} - \frac{\partial}{\partial t}\int \frac{\partial \rho_e^{(2)}(\vec{x},t)}{\partial t}\chi(\vec{x},t)d\vec{x} + \int \frac{\partial^2 \rho_e^{(2)}(\vec{x},t)}{\partial t^2}\chi(\vec{x},t)d\vec{x}$$
$$\tag{30}$$

Rearrange terms to obtain

$$\frac{d\xi_o^{(2)}}{dt} = \int \chi(\vec{x},t)\frac{\partial}{\partial t}\left( \frac{\partial \rho_e^{(2)}(\vec{x},t)}{\partial t} + \vec{\nabla} \cdot \vec{J}_e^{(2)}(\vec{x},t) \right)d\vec{x} - \frac{\partial}{\partial t}\int \frac{\partial \rho_e^{(2)}(\vec{x},t)}{\partial t}\chi(\vec{x},t)d\vec{x} \tag{31}$$

Use this result in the expression

$$\xi_o^{(2)}(t_b) = \xi_o^{(2)}(t_a) + \int\limits_{t_a}^{t_b} \frac{d\xi_o^{(2)}(t)}{dt} dt \tag{32}$$

to obtain

$$\xi_o^{(2)}(t_b) = \xi_o^{(2)}(t_a) - \int \frac{\partial \rho_e^{(2)}(\vec{x}, t_b)}{\partial t} \chi(\vec{x}, t_b) d\vec{x}$$
$$+ \int\limits_{t_a}^{t_b} dt \int \chi(\vec{x}, t) \frac{\partial}{\partial t}\left(\frac{\partial \rho_e^{(2)}(\vec{x}, t)}{\partial t} + \vec{\nabla} \cdot \vec{J}_e^{(2)}(\vec{x}, t)\right) d\vec{x} \tag{33}$$

where we have used the fact that $\chi(\vec{x}, t_a) = 0$ (see Eq. (24)).

Use Eqs. (25), (26), and (27) in (33) to obtain

$$\xi_o^{(2)}(t_b) = \xi_o^{(1)}(t_b) - \int \frac{\partial \rho_e^{(1)}(\vec{x}, t_b)}{\partial t} \chi(\vec{x}, t_b) d\vec{x}$$
$$+ \int\limits_{t_a}^{t_b} dt \int \chi(\vec{x}, t) \frac{\partial L^{(1)}(\vec{x}, t)}{\partial t} d\vec{x} \tag{34}$$

where

$$L^{(1)}(\vec{x}, t) \equiv \frac{\partial \rho_e^{(1)}(\vec{x}, t)}{\partial t} + \vec{\nabla} \cdot \vec{J}_e^{(1)}(\vec{x}, t) \tag{35}$$

In Eq. (34) the quantities $\xi_o^{(1)}(t_b)$, $\rho_e^{(1)}(\vec{x}, t_b)$ and $L^{(1)}(\vec{x}, t)$ are independent of $\chi$.

Note that

$$L^{(1)}(\vec{x}, t) = 0 \tag{36}$$

is the continuity equation. Since charge conservation is an established experimental fact we could at this point use Eq. (36) to simplify (34). However, the proof is not dependent on the truth of the continuity equation, only on gauge invariance. Therefore we shall consider two possibilities.

In one case assume that Eq. (36) is true or more specifically that

$$\frac{\partial L^{(1)}(\vec{x}, t)}{\partial t} = 0 \tag{37}$$

Use this in Eq. (34) to obtain

$$\xi_o^{(2)}(t_b) = \xi_o^{(1)}(t_b) - \int \frac{\partial \rho_e^{(1)}(\vec{x}, t_b)}{\partial t} \chi(\vec{x}, t_b) d\vec{x} \tag{38}$$

Now it is possible to find a $\chi(\vec{x}, t)$ such that $\xi_o^{(2)}(t_b)$ is a negative number with an arbitrarily large magnitude. For example let

$$\chi(\vec{x}, t_b) = f \frac{\partial \rho_e^{(1)}(\vec{x}, t_b)}{\partial t} \tag{39}$$

where f is a constant. Other then this $\chi(\vec{x}, t)$ is arbitrary (except that it must also satisfy the initial conditions in Eq. (24)). Use Eq. (39) in (38) to obtain

$$\xi_o^{(2)}(t_b) = \xi_o^{(1)}(t_b) - f \int \left( \frac{\partial \rho_e^{(1)}(\vec{x}, t_b)}{\partial t} \right)^2 d\vec{x} \tag{40}$$

The quantity under the integral sign is always positive. Therefore, given Eq. (21), the integral is greater than zero so as $f \to \infty$ then $\xi_o^{(2)}(t_b) \to -\infty$.

Next consider the possibility that

$$\frac{\partial L^{(1)}(\vec{x}, t)}{\partial t} \neq 0 \tag{41}$$

Then we can set

$$\chi(\vec{x}, t) = \left\langle \begin{array}{l} -f \dfrac{\partial L^{(1)}(\vec{x}, t)}{\partial t} \text{ for } t_a < t < t_b \\ \quad 0 \text{ for } t \leq t_a \text{ or } t \geq t_b \end{array} \right. \tag{42}$$

Use this in Eq. (34) to yield

$$\xi_o^{(2)}(t_b) = \xi_o^{(1)}(t_b) - f \int\limits_{t_a}^{t_b} dt \int \left\{ \frac{\partial L^{(1)}(\vec{x}, t)}{\partial t} \right\}^2 d\vec{x}$$

Once again the integral is always positive so that as $f \rightarrow \infty$ then $\xi_o^{(2)}(t_b) \rightarrow -\infty$.

<u>IV Discussion</u>

Let us review the results of the previous section. We start out with a quantum state $\left| \Omega_1(t) \right\rangle$ such that the condition given by Eq. (21) is true at some point in time $t=t_b$. $\left| \Omega_1(t) \right\rangle$ satisfies Eq. (1) where the electric potential is zero. Then at some time $t=t_a<t_b$ define the quantum state $\left| \Omega_2(t_a) \right\rangle = \left| \Omega_1(t_a) \right\rangle$. Next, use Eq. (1) to evolve $\left| \Omega_2(t) \right\rangle$ forward in time to $t_b$ in the presence of an electric potential that is related to the electric potential of $\left| \Omega_1(t) \right\rangle$ by a gauge transformation. These leads to Eq. (34) for the free field energy of $\left| \Omega_2(t) \right\rangle$ at $t=t_b$. In this expression the quantities $\vec{J}_e^{(1)}$ and $\rho_e^{(1)}$ are independent of $\chi$. Therefore $\chi$ may be defined as discussed above to make $\xi_o^{(2)}(t_b)$ less than $\xi_o^{(1)}(t_b)$ by an arbitrarily large amount. Therefore, there is in principle, no lower bound to the free field energy for Dirac field theory.

Now this may seem like a surprising result because the generally held belief is that the vacuum state is the minimum value of the free field energy (see Chap. 9 of Greiner et al[4] or Chap. 1 of Pauli[3]). What has been shown here is that there must exist quantum states whose free field energy is less than that of the vacuum state. In the next section we will discuss the properties of the vacuum state as they are normally defined in quantum field theory and then, in the following section, it will be shown how the vacuum state can be redefined to be consistent with these results.

<div align="center"><u>V. The vacuum state</u></div>

Define the field operator in the form of the following expansion,

$$\hat{\psi}(\vec{x}) = \sum_n \hat{a}_n \phi_n(\vec{x}); \quad \hat{\psi}^\dagger(\vec{x}) = \sum_n \hat{a}_n^\dagger \phi_n^\dagger(\vec{x}) \tag{43}$$

where the $\hat{a}_n$ ($\hat{a}_n^\dagger$) are the destruction(creation) operators for a particle in the state $\phi_n(\vec{x})$. They satisfy the anticommutator relation

$$\hat{a}_m \hat{a}_n^\dagger + \hat{a}_n^\dagger \hat{a}_m = \delta_{mn}; \text{ all other anticommutators} = 0 \tag{44}$$

The $\phi_n(\vec{x})$ are eigenfunctions of the free field single particle Dirac equation with energy eigenfunction $\lambda_n E_n$. That is,

$$\lambda_n E_n \phi_n(\vec{x}) = H_o \phi_n(\vec{x}) \tag{45}$$

where

$$H_o = -i\vec{\alpha} \cdot \vec{\nabla} + \beta m \tag{46}$$

and where

$$E_n = +\sqrt{\vec{p}_n^2 + m^2}, \quad \lambda_n = \begin{cases} +1 \text{ for a positive energy state} \\ -1 \text{ for a negative energy state} \end{cases} \tag{47}$$

where $\vec{p}_n$ is the momentum of the state n. Solutions of (45) are of the form

$$\phi_n(\vec{x}) = u_n e^{i\vec{p}_n \cdot \vec{x}} \tag{48}$$

where $u_n$ is a constant 4-spinor. The $\phi_n(\vec{x})$ form a complete orthonormal basis in Hilbert space and satisfy

$$\sum_n \left(\phi_n^\dagger(\vec{x})\right)_{(a)} \left(\phi_n(\vec{y})\right)_{(b)} = \delta_{ab}\delta^3(\vec{x} - \vec{y}) \qquad (49)$$

and

$$\int \phi_n^\dagger(\vec{x})\phi_m(\vec{x})d\vec{x} = \delta_{mn} \qquad (50)$$

where "a" and "b" are spinor indices (see page 107 of Heitler [5]).

Following Greiner [4] define the state vector $\left|0, \text{bare}\right\rangle$ which is the state vector that is empty of all particles, i.e.,

$$\hat{a}_n \left|0, \text{bare}\right\rangle = 0 \text{ for all n} \qquad (51)$$

For the index 'n' we will define the following

$$n<0 \text{ refers to negative energy states.} \qquad (52)$$

$$n>0 \text{ refers to positive energy states.}$$

The vacuum state vector $|0\rangle$ is defined as the state vector in which all negative energy states are occupied by a single particle. Therefore

$$|0\rangle = \prod_{n<0} \hat{a}_n^\dagger |0, \text{bare}\rangle \qquad (53)$$

where, as defined above, n<0 means that the product is taken over all negative energy states. From this expression, and Eqs. (44) and (51), $|0\rangle$ can then be defined by

$$\hat{a}_n|0\rangle = 0 \text{ for } n > 0 \text{ ; } \hat{a}_n^\dagger|0\rangle = 0 \text{ for } n < 0 \qquad (54)$$

(Note, it is possible, for n<0, to replace the electron destruction and creation operators, $\hat{a}_n$ and $\hat{a}_n^\dagger$, with the positron creation and destruction operators, $\hat{b}_n^\dagger$ and $\hat{b}_n$, respectively. However, in this article, it will be convenient to stick with the present notation).

Next define the operators $\hat{H}_o, \hat{\vec{J}}$, and $\hat{\rho}$ in terms of the field operators. There are a number of ways of preceding. For example $\hat{\rho}$ can be expressed as $q\left[\hat{\psi}^\dagger(\vec{x}), \hat{\psi}(\vec{x})\right]$ or $q{:}\hat{\psi}^\dagger(\vec{x})\hat{\psi}(\vec{x}){:}$ where q is the electric charge. All these expressions are basically equivalent and simply represent different ways of subtracting out the infinite vacuum charge. In this discussion it will be convenient to use

$$\hat{\rho} = q\hat{\psi}^\dagger\hat{\psi} - \rho_R \tag{55}$$

$$\hat{\vec{J}} = q\hat{\psi}^\dagger\vec{\alpha}\hat{\psi} - \vec{J}_R \tag{56}$$

$$\hat{H}_o = \int \hat{\psi}^\dagger H_o \hat{\psi} d\vec{x} - \xi_R \tag{57}$$

Where the renormalization constant $\rho_R, \vec{J}_R$, and $\xi_R$ are used to make the vacuum expectation values of the above operators equal to zero.

Any arbitrary state $|\Omega\rangle$ can be expanded in terms of a set of orthonormal basis states $|\varphi_n\rangle$ which are created by the action of the operators $\hat{a}_n^\dagger$ and $a_n$ on $|0\rangle$ (see Chapt. 3 of Itzykson and Zuber [6]). Therefore $|\Omega\rangle$ can be written as

$$|\Omega\rangle = \sum_r c_r |\varphi_r\rangle \tag{58}$$

where the $c_r$ are expansion coefficients and where

$$\langle\varphi_m|\varphi_n\rangle = \delta_{nm} \tag{59}$$

The basis states $|\varphi_n\rangle$ are eigenstates of $\hat{H}_o$ ,i.e.,

$$\hat{H}_o|\varphi_n\rangle = \varepsilon_n|\varphi_n\rangle \tag{60}$$

The eigenvalues $\varepsilon_n$ are all positive (or zero for the vacuum state). This is because the action of the operators $\hat{a}_n^\dagger$ and $a_n$ on $|0\rangle$ is to either create a previously unoccupied positive energy state or to destroy an occupied negative energy state, respectively. If n=0 refers to the vacuum state we have that

$$\varepsilon_n > \varepsilon_o = 0 \ \text{ for n} \neq 0 \tag{61}$$

Using Eqs. (60) and (59) in (3) we have that the free field energy for a normalized quantum state $|\Omega\rangle$ is

$$\xi_o\left(|\Omega\rangle\right) = \sum_m |c_m|^2 \varepsilon_m \tag{62}$$

Use Eq. ( 61) in the above to yield

$$\xi_o\left(|\Omega\rangle\right) > \xi_o\left(|0\rangle\right) = 0 \ \text{ if } |\Omega\rangle \neq |0\rangle \tag{63}$$

This equation states that the free field energy of the vacuum state is lower than that of any other state. This is because, as just discussed, any arbitrary state $|\Omega\rangle$ can be expressed as a combination of the state $|0\rangle$ and the other basis states $|\varphi_m\rangle$ which are formed from the action of the operators $\hat{a}_n^\dagger$ or $a_n$ acting on $|0\rangle$. These other basis states have free field energy greater than that of $|0\rangle$. Therefore the free field energy of $|\Omega\rangle$ is always greater then that of $|0\rangle$ (unless $|\Omega\rangle = |0\rangle$).

## VI. Redefining the Vacuum state

In Section III it was shown that gauge invariance requires that quantum states must exist whose free field energy is less than that of the vacuum state $|0\rangle$. However when $|0\rangle$ is defined as in the previous section it is seen that $|0\rangle$ has the lowest energy free field of any state. Hence we have an inconsistency. Dirac quantum field theory cannot be both gauge invariant and have $|0\rangle$ as the vacuum state.

In the following it will be shown how this inconsistency can be eliminated by a modest modification to the definition of the vacuum state. The immediate problem with $|0\rangle$ is that the basis states that are derived from $|0\rangle$ are all positive energy states. As we have seen this means the free field energy of any arbitrary state must be positive. Therefore a minimum requirement for a consistent theory is that basis states must exist with less energy then the vacuum state.

To see how this can be done, start by defining the state vector $|0, \Delta E_w\rangle$ as follows. The top of the negative energy band has an energy of $-m$. Define $|0, \Delta E_w\rangle$ as the multiparticle quantum state in which each single particle state in the band of negative energy states from energy $-m$ to $-(m+\Delta E_w)$ is occupied and all other single particle states are unoccupied. Let the notation "$n \in$ band" mean that "$n$" is a single particle quantum state with energy in the range $-m$ to $-(m+\Delta E_w)$. Then $|0, \Delta E_w\rangle$ can be defined by

$$|0, \Delta E_w\rangle \equiv \prod_{n \in \text{band}} \hat{a}_n^\dagger |0, \text{bare}\rangle \qquad (64)$$

Let $n <$ band refer to the single particle quantum state with energy less than $-(m+\Delta E_w)$. Recall, also, that $n > 0$ refers to positive energy states. Therefore $|0, \Delta E_w\rangle$ can be defined by

$$\hat{a}_n |0, \Delta E_w\rangle = 0 \text{ for } n > 0$$
$$\hat{a}_n^\dagger |0, \Delta E_w\rangle = 0 \text{ for } n \in \text{band} \qquad (65)$$
$$\hat{a}_n |0, \Delta E_w\rangle = 0 \text{ for } n < \text{band}$$

$|0, \Delta E_w \rangle$ is an eigenvector of the operator $\hat{H}_o$. The eigenvalue can be set equal to zero by proper selection of the constant $\xi_R$ in Eqs.(57) therefore we write,

$$\hat{H}_o \big| 0, \Delta E_w \big\rangle = 0 \tag{66}$$

We redefine the vacuum state as follows. Let the vacuum be the state $\big| 0, \Delta E_w \to \infty \big\rangle$ which is the state $\big| 0, \Delta E_w \big\rangle$ in the limit that $\Delta E_w \to \infty$. In calculations involving $\big| 0, \Delta E_w \to \infty \big\rangle$, $\Delta E_w$ is assumed to be finite with the limit $\Delta E_w \to \infty$ taken at the end of the calculation.

It has been shown in Section III that quantum states exist whose free field energy is less than that of the vacuum state. This is not possible if $|0\rangle$ is the vacuum state. But using $|0, \Delta E_w{\to}\infty\rangle$ as the vacuum state allows this because then basis states will exist that have less energy than the vacuum state. Let $m \in$ band and $n <$ band. The operator pair $\hat{a}_n^\dagger \hat{a}_m$ acting on $|0, \Delta E_w{\to}\infty\rangle$ will destroy a particle with the quantum number 'm' within the band and produce a particle with quantum number 'n' underneath the band. This will produce a quantum state with energy less then $|0, \Delta E_w{\to}\infty\rangle$ by the amount $E_n - E_m$ (note that $E_n > E_m$ because 'n' is a single particle quantum state with less energy (more negative) than 'm'). An arbitrary state can, then, be expanded in terms of basis states whose free field energy can be greater than or less than that of the vacuum. Therefore the free field energy of an arbitrary state can be less than the free field energy of the redefined vacuum state $|0, \Delta E_w{\to}\infty\rangle$. Therefore the quantum state $|0, \Delta E_w{\to}\infty\rangle$ is consistent with the requirements of gauge invariance.

One problem that must be addressed is does the state $|0, \Delta E_w \to \infty\rangle$ give stability to positive energy particles? An important role that the vacuum state $|0\rangle$ plays is to prevent the scattering of positive energy particles into unoccupied negative energy states by a perturbing potential. Does $|0, \Delta E_w \to \infty\rangle$ allow positive energy particles to scatter into unoccupied states that exist underneath the bottom edge of the negative energy band? A strong argument that $|0, \Delta E_w \to \infty\rangle$ provides the stability needed is to note that in perturbation theory terms of the following form will appear:

$$\int \phi_n^\dagger(\vec{x}, t) V(\vec{x}, t) \Phi(\vec{x}, t) d\vec{x} dt \tag{67}$$

where $\Phi(\vec{x}, t)$ is some positive energy wave function, $V(\vec{x}, t)$ is some perturbing potential, and $\phi_n(\vec{x}, t)$ is the unoccupied negative energy state underneath the negative band for which $|\vec{p}_n| \to \infty$. Now

$$\phi_n(\vec{x}, t) = u_n e^{i\vec{p}_n \cdot \vec{x}} e^{-i\lambda_n E_n t} \tag{68}$$

Since $|\vec{p}_n| \to \infty$ and $E_n \to \infty$ the wave function $\phi_n(\vec{x}, t)$ oscillates at a rate approaching infinity in both space and time. If we assume that the oscillatory behavior of the positive energy wave function $\Phi(\vec{x}, t)$ and the perturbing potential $V(\vec{x}, t)$ is finite then the integrand is dominated by the rapid oscillation of $\phi_n(\vec{x}, t)$ and the integral will approach zero as $|\vec{p}_n| \to \infty$. Therefore the transition probability from the positive energy wave function into states underneath the negative energy band will become arbitrarily small as $\Delta E_w \to \infty$.

Another feature of $|0, \Delta E_w \to \infty\rangle$ is that it is consistent with positron theory. In this regard it is identical to the state $|0\rangle$ in that a "hole" will look like a positron. Therefore $|0, \Delta E_w \rightarrow \infty\rangle$ meets all requirement for a vacuum state. It stabilizes positive energy particles against transitions to negative energy states, is consistent with positron theory, and is consistent with the requirements of gauge invariance.

### VII. The vacuum state and gauge invariance

In this section it will be shown by direct calculation that using the state $|0\rangle$ as the vacuum state will destroy the gauge invariance of the theory. Referring to Eq. (1) we have that

$$\left|\Omega(t+\Delta t)\right\rangle \underset{\Delta t \rightarrow 0}{=} \left|\Omega(t)\right\rangle - i\hat{H}(t)\left|\Omega(t)\right\rangle \Delta t + O\left(\Delta t^2\right) \tag{69}$$

$$\left\langle \Omega(t+\Delta t)\right|_{\Delta t \rightarrow 0} = \left\langle \Omega(t)\right| + i\left\langle \Omega(t)\right|\hat{H}(t)\Delta t + O\left(\Delta t^2\right) \tag{70}$$

Use this in Eq. (6) to obtain

$$\vec{J}_e(\vec{x}, t+\Delta t) = \left\langle \Omega(t+\Delta t)\left|\hat{\vec{J}}(\vec{x})\right|\Omega(t+\Delta t)\right\rangle$$
$$\underset{\Delta t \rightarrow 0}{=} \left\langle \Omega(t)\left|\hat{\vec{J}}(\vec{x})\right|\Omega(t)\right\rangle + i\left\langle \Omega(t)\left|\left[\hat{H}(t), \hat{\vec{J}}(\vec{x})\right]\right|\Omega(t)\right\rangle \Delta t + O\left(\Delta t^2\right) \tag{71}$$

Let the electric potential be given by

$$\left(A_o(\vec{x}, t), \vec{A}(\vec{x}, t)\right) = \left(\frac{\partial \chi(\vec{x}, t)}{\partial t}, \left(-\vec{\nabla}\chi(\vec{x}, t)\right)\right) \tag{72}$$

Use this and Eq. (2) in (71) to obtain

$$\vec{J}_e(\vec{x}, t+\Delta t) \underset{\Delta t \rightarrow 0}{=} \vec{J}_e(\vec{x}, t) + i\left\langle \Omega(t)\left|\left[\hat{H}_o(t), \hat{\vec{J}}(\vec{x})\right]\right|\Omega(t)\right\rangle \Delta t$$
$$+ i\left\langle \Omega(t)\left|\left[\left(\int \hat{\vec{J}}(\vec{y}) \cdot \vec{\nabla}\chi(\vec{y}, t)d\vec{y} + \int \hat{\rho}(\vec{y})\frac{\partial \chi(\vec{y}, t)}{\partial t}d\vec{y}\right), \hat{\vec{J}}(\vec{x})\right]\right|\Omega(t)\right\rangle \Delta t + O\left(\Delta t^2\right) \tag{73}$$

Rearrange terms to obtain

$$\vec{J}_e(\vec{x}, t + \Delta t) \underset{\Delta t \to 0}{=} \vec{J}_e(\vec{x}, t) + i \langle \Omega(t) | \left[ \hat{H}_o(t), \hat{\vec{J}}(\vec{x}) \right] | \Omega(t) \rangle \Delta t$$

$$+ i \Delta t \int \left\{ \langle \Omega(t) | \left[ \hat{\vec{J}}(\vec{y}), \hat{\vec{J}}(\vec{x}) \right] | \Omega(t) \rangle \cdot \vec{\nabla} \chi(\vec{y}, t) + \langle \Omega(t) | \left[ \hat{\rho}(\vec{y}), \hat{\vec{J}}(\vec{x}) \right] | \Omega(t) \rangle \frac{\partial \chi(\vec{y}, t)}{\partial t} \right\} d\vec{y} + O\left( \Delta t^2 \right)$$

$$(74)$$

If quantum theory is gauge invariant then $\vec{J}_e(\vec{x}, t + \Delta t)$ must not be dependent on $\chi$ because the electric potential of Eq. (72) is a gauge transformation from the case where the electric potential is zero. Therefore, since $\frac{\partial \chi}{\partial t}$ is arbitrary, the quantity $\langle \Omega(t) | \left[ \hat{\rho}(\vec{y}), \hat{\vec{J}}(\vec{x}) \right] | \Omega(t) \rangle$ must equal zero for any $|\Omega(t)\rangle$. Check the case for which $|\Omega(t)\rangle = |0\rangle$. Define

$$\vec{I}(\vec{x}, \vec{y}) \equiv \langle 0 | \left[ \hat{\rho}(\vec{y}), \hat{\vec{J}}(\vec{x}) \right] | 0 \rangle = \langle 0 | \hat{\rho}(\vec{y}) \hat{\vec{J}}(\vec{x}) | 0 \rangle - \langle 0 | \hat{\vec{J}}(\vec{x}) \hat{\rho}(\vec{y}) | 0 \rangle \qquad (75)$$

$\vec{I}(\vec{x}, \vec{y})$ is called the Schwinger term. If quantum field theory is gauge invariant then the Schwinger term $\vec{I}(\vec{x}, \vec{y})$ must equal zero. Use (43) in (55) and (56) to obtain

$$\hat{\rho}(\vec{y}) = q \sum_{n,m} \hat{a}_n^\dagger \hat{a}_m \phi_n^\dagger(\vec{y}) \phi_m(\vec{y}) - \rho_R \qquad (76)$$

and

$$\hat{\vec{J}}(\vec{x}) = q \sum_{n,m} \hat{a}_n^\dagger \hat{a}_m \phi_n^\dagger(\vec{x}) \vec{\alpha} \phi_m(\vec{x}) - \vec{J}_R \qquad (77)$$

From the definition of the vacuum state (Eq. (55) and (56)) and the anticommutator relationships (Eq. (44)) we have that

$$\hat{\vec{J}}(\vec{x})|0\rangle = q \sum_{\substack{n>0 \\ m<0}} \hat{a}_n^\dagger \hat{a}_m \phi_n^\dagger(\vec{x})\vec{\alpha}\phi_m(\vec{x})|0\rangle - \vec{J}_1|0\rangle \tag{78}$$

where $\vec{J}_1$ is given by

$$\vec{J}_1 = \vec{J}_R - q \sum_{n<0} \phi_n^\dagger(\vec{x})\vec{\alpha}\phi_n(\vec{x}) \tag{79}$$

From Eq. (48) it is seen that $\vec{J}_1$ is constant. Similarly

$$\langle 0|\hat{\rho}(\vec{y}) = q\langle 0| \sum_{\substack{n>0 \\ m<0}} \hat{a}_m^\dagger \hat{a}_n \phi_m^\dagger(\vec{y})\phi_n(\vec{y}) - \langle 0|\rho_1 \tag{80}$$

where $\rho_1$ is a constant given by

$$\rho_1 = \rho_R - q \sum_{n<0} \phi_n^\dagger(\vec{y})\phi_n(\vec{y}) \tag{81}$$

Use Eqs. (78) and (80) in (75) to obtain

$$\vec{I}(\vec{x},\vec{y}) = \left( q^2 \sum_{\substack{n>0 \\ m<0}} \left(\phi_m^\dagger(\vec{y})\phi_n(\vec{y})\right)\left(\phi_n^\dagger(\vec{x})\vec{\alpha}\phi_m(\vec{x})\right) \right) - (\text{h.c.}) \tag{82}$$

where (h.c.) means to take the hermitian conjugate of the preceding term.

Now if $\vec{I}(\vec{x},\vec{y}) = 0$ then $\vec{\nabla}_{\vec{x}} \cdot \vec{I}(\vec{x},\vec{y}) = 0$. From the above equation we obtain

$$\vec{\nabla}_{\vec{x}} \cdot \vec{I}(\vec{x},\vec{y}) = \left( q^2 \sum_{\substack{n>0 \\ m<0}} \phi_m^\dagger(\vec{y})\phi_n(\vec{y})\vec{\nabla}\cdot\left(\phi_n^\dagger(\vec{x})\vec{\alpha}\phi_m(\vec{x})\right) \right) - (\text{h.c.}) \tag{83}$$

To evaluate this expression use the following,

$$\vec{\nabla}\cdot\left(\phi_n^\dagger(\vec{x})\vec{\alpha}\phi_m(\vec{x})\right) = \left(\vec{\alpha}\cdot\vec{\nabla}\phi_n(\vec{x})\right)^\dagger \phi_m(\vec{x}) + \phi_n^\dagger(\vec{x})\vec{\alpha}\cdot\vec{\nabla}\phi_m(\vec{x}) \tag{84}$$

Next, from the definition of $H_o$ (Eq. (46)), we obtain

$$\vec{\nabla} \cdot \left( \phi_n^\dagger(\vec{x}) \vec{\alpha} \phi_m(\vec{x}) \right) = -i \left\{ \left( H_o \phi_n(\vec{x}) \right)^\dagger \phi_m(\vec{x}) - \phi_n^\dagger(\vec{x}) H_o \phi_m(\vec{x}) \right\} \tag{85}$$

Now use Eq. (45) in the above to obtain

$$\vec{\nabla} \cdot \left( \phi_n^\dagger(\vec{x}) \vec{\alpha} \phi_m(\vec{x}) \right) = -i \left( \lambda_n E_n - \lambda_m E_m \right) \phi_n^\dagger(\vec{x}) \phi_m(\vec{x}) \tag{86}$$

Use this in Eq. (83) to obtain

$$\vec{\nabla}_{\vec{x}} \cdot \vec{I}(\vec{x}, \vec{y}) = \left( -iq^2 \sum_{\substack{n>0 \\ m<0}} \phi_m^\dagger(\vec{y}) \phi_n(\vec{y}) \left( \lambda_n E_n - \lambda_m E_m \right) \phi_n^\dagger(\vec{x}) \phi_m(\vec{x}) \right) - (\text{h.c.}) \tag{87}$$

Use Eq. (47) to obtain

$$\vec{\nabla}_{\vec{x}} \cdot \vec{I}(\vec{x}, \vec{y}) = \left( -iq^2 \sum_{\substack{n>0 \\ m<0}} \left( E_n + E_m \right) \phi_m^\dagger(\vec{y}) \phi_n(\vec{y}) \phi_n^\dagger(\vec{x}) \phi_m(\vec{x}) \right) - (\text{h.c.}) \tag{88}$$

Evaluate this at $\vec{y} = \vec{x}$ to obtain

$$\vec{\nabla}_{\vec{x}} \cdot \vec{I}(\vec{x}, \vec{y}) \Big|_{\vec{y}=\vec{x}} = \left( -iq^2 \sum_{\substack{n>0 \\ m<0}} \left( E_n + E_m \right) \phi_m^\dagger(\vec{x}) \phi_n(\vec{x}) \phi_n^\dagger(\vec{x}) \phi_m(\vec{x}) \right) - (\text{h.c.}) \tag{89}$$

This yields

$$\vec{\nabla}_{\vec{x}} \cdot \vec{I}(\vec{x}, \vec{y}) \Big|_{\vec{y}=\vec{x}} = \left( -iq^2 \sum_{\substack{n>0 \\ m<0}} \left( E_n + E_m \right) \left| \phi_m^\dagger(\vec{x}) \phi_n(\vec{x}) \right|^2 \right) - (\text{h.c.})$$

$$= -2i \sum_{\substack{n>0 \\ m<0}} q^2 \left( E_n + E_m \right) \left| \phi_m^\dagger(\vec{x}) \phi_n(\vec{x}) \right|^2 \tag{90}$$

Each term in the sum is positive therefore the above expression is not equal to zero.

Therefore $\vec{I}(\vec{x}, \vec{y}) \neq 0$ so that the theory is not gauge invariant when the vacuum state is

$\left| 0 \right\rangle$.

Now let us work the same problem using the quantum state $\left| 0, \Delta E_w \right\rangle$. Define

$$\vec{I}'(\vec{x}, \vec{y}) \equiv \left\langle 0, \Delta E_w \left| \left[ \hat{\rho}(\vec{y}), \hat{\vec{J}}(\vec{x}) \right] \right| 0, \Delta E_w \right\rangle \tag{91}$$

From Eqs. (56) and (65) we obtain

$$
\begin{aligned}
\hat{\vec{J}}(\vec{x}) \left| 0, \Delta E_w \right\rangle =& \sum_{\substack{m \in \text{band} \\ n > 0}} q \phi_n^\dagger(\vec{x}) \vec{\alpha} \phi_m(\vec{x}) \hat{a}_n^\dagger \hat{a}_m \left| 0, \Delta E_w \right\rangle \\
&+ \sum_{\substack{m \in \text{band} \\ n < \text{band}}} q \phi_n^\dagger(\vec{x}) \vec{\alpha} \phi_m(\vec{x}) \hat{a}_n^\dagger \hat{a}_m \left| 0, \Delta E_w \right\rangle - \vec{J}_1' \left| 0, \Delta E_w \right\rangle
\end{aligned} \tag{92}
$$

where $\vec{J}_1'$ is a constant given by

$$\vec{J}_1' = \vec{J}_R - \sum_{m \in \text{band}} q \phi_m^\dagger(\vec{x}) \vec{\alpha} \phi_m(\vec{x}) \tag{93}$$

Similarly

$$
\begin{aligned}
\left\langle 0, \Delta E_w \left| \hat{\rho}(\vec{y}) \right. = \right. & \left\langle 0, \Delta E_w \left| \sum_{\substack{m \in \text{band} \\ n > 0}} q \hat{a}_m^\dagger \hat{a}_n \phi_m^\dagger(\vec{y}) \phi_n(\vec{y}) \right. \right. \\
&+ \left\langle 0, \Delta E_w \left| \sum_{\substack{m \in \text{band} \\ n < \text{band}}} q \hat{a}_m^\dagger \hat{a}_n \phi_m^\dagger(\vec{y}) \phi_n(\vec{y}) - \left\langle 0, \Delta E_w \left| \rho_1' \right. \right. \right. \right.
\end{aligned} \tag{94}
$$

where $\rho_1'$ is a constant given by

$$\rho_1' = \rho_R - \sum_{m \in \text{band}} q \phi_m^\dagger \phi_m \tag{95}$$

Therefore

$$\vec{I}'(\vec{x}, \vec{y}) = q^2 \left\{ \sum_{\substack{m \in band \\ n > 0}} \phi_m^\dagger(\vec{y})\phi_n(\vec{y})\phi_n^\dagger(\vec{x})\vec{\alpha}\phi_m(\vec{x}) + \sum_{\substack{m \in band \\ n < band}} \phi_m^\dagger(\vec{y})\phi_n(\vec{y})\phi_n^\dagger(\vec{x})\vec{\alpha}\phi_m(\vec{x}) \right\} - \{\text{h.c.}\}$$

$$(96)$$

Next define

$$F_2 \equiv \sum_{\substack{m \in band \\ n \in band}} \phi_m^\dagger(\vec{y})\phi_n(\vec{y})\phi_n^\dagger(\vec{x})\vec{\alpha}\phi_m(\vec{x}) \tag{97}$$

Therefore

$$F_2^\dagger \equiv \sum_{\substack{m \in band \\ n \in band}} \phi_n^\dagger(\vec{y})\phi_m(\vec{y})\phi_m^\dagger(\vec{x})\vec{\alpha}\phi_n(\vec{x}) \tag{98}$$

One can interchange the dummy indices n and m in Eq. (97) to obtain

$$F_2 = F_2^\dagger \Rightarrow F_2 - F_2^\dagger = 0 \tag{99}$$

Therefore the quantity $q^2\left(F_2 - F_2^\dagger\right) = 0$ can be added to the right side of Eq. (96) to

obtain

$$\vec{I}'(\vec{x}, \vec{y}) = q^2 \left\{ \sum_{\substack{m \in band \\ n > 0}} \phi_m^\dagger(\vec{y})\phi_n(\vec{y})\phi_n^\dagger(\vec{x})\vec{\alpha}\phi_m(\vec{x}) + F_2 + \sum_{\substack{m \in band \\ n < band}} \phi_m^\dagger(\vec{y})\phi_n(\vec{y})\phi_n^\dagger(\vec{x})\vec{\alpha}\phi_m(\vec{x}) \right\} - \{\text{h.c.}\}$$

$$(100)$$

Use the expression for $F_2$ from Eq. (97) in the above expression to obtain

$$\vec{I}'(\vec{x}, \vec{y}) = q^2 \left\{ \sum_{\substack{m \in band \\ n > 0}} \phi_m^\dagger(\vec{y}) \phi_n(\vec{y}) \phi_n^\dagger(\vec{x}) \vec{\alpha} \phi_m(\vec{x}) + \sum_{\substack{m \in band \\ n \in band}} \phi_m^\dagger(\vec{y}) \phi_n(\vec{y}) \phi_n^\dagger(\vec{x}) \vec{\alpha} \phi_m(\vec{x}) \right. \\ \left. + \sum_{\substack{m \in band \\ n < band}} \phi_m^\dagger(\vec{y}) \phi_n(\vec{y}) \phi_n^\dagger(\vec{x}) \vec{\alpha} \phi_m(\vec{x}) \right\} - \{h.c.\}$$

(101)

Next use

$$\sum_{\substack{m \in band \\ all\ n}} = \sum_{\substack{m \in band \\ n > 0}} + \sum_{\substack{m \in band \\ n \in band}} + \sum_{\substack{m \in band \\ n < band}}$$

(102)

in the above expression to yield

$$\vec{I}'(\vec{x}, \vec{y}) = q^2 \left\{ \sum_{\substack{m \in band \\ all\ n}} \phi_m^\dagger(\vec{y}) \phi_n(\vec{y}) \phi_n^\dagger(\vec{x}) \vec{\alpha} \phi_m(\vec{x}) \right\} - \{h.c.\}$$

(103)

In the above expression do the summation over the index 'n' first and use Eq. (49) to yield

$$\vec{I}'(\vec{x}, \vec{y}) = \left\{ q^2 \sum_{m \in band} \phi_m^\dagger(\vec{y}) \vec{\alpha} \phi_m(\vec{x}) \delta^3(\vec{x} - \vec{y}) \right\} - \{h.c.\}$$

(104)

Now use the relationship

$$f(\vec{y}) \delta^3(\vec{x} - \vec{y}) = f(\vec{x})$$

(105)

where $f(\vec{y})$ is an arbitrary function to obtain

$$\vec{I}'(\vec{x}, \vec{y}) = \left\{ q^2 \sum_{m \in band} \phi_m^\dagger(\vec{x}) \vec{\alpha} \phi_m(\vec{x}) \delta^3(\vec{x} - \vec{y}) \right\} - \{h.c.\}$$

(106)

Use the fact that

$$\left( \phi_m^\dagger(\vec{x}) \vec{\alpha} \phi_m(\vec{x}) \right)^\dagger = \phi_m^\dagger(\vec{x}) \vec{\alpha} \phi_m(\vec{x})$$

(107)

to obtain

$$\vec{I}'(\vec{x}, \vec{y}) = 0 \tag{108}$$

Therefore the theory is gauge invariant if $\left|0, \Delta E_w \to \infty\right\rangle$ is used as the vacuum state.

Now what is the difference between $\left|0\right\rangle$ and $\left|0, \Delta E_w \to \infty\right\rangle$ that accounts for the fact that $\vec{I}(\vec{x}, \vec{y}) \neq 0$ but that $\vec{I}'(\vec{x}, \vec{y}) = 0$. The key difference can be seen by comparing

Eq. (82) and (96). In the expression for $\vec{I}(\vec{x}, \vec{y})$ (Eq. (82)) the only transitions are from occupied negative energy states to unoccupied positive energy states. For $\vec{I}'(\vec{x}, \vec{y})$ (Eq.

(96)) there is an analogous contribution due to transitions from occupied states within the negative energy band to unoccupied positive energy states. However, there is also an additional term due to transitions from states within the band to unoccupied negative energy states underneath the band. It is the contribution made by this additional term that makes $\vec{I}'(\vec{x}, \vec{y}) = 0$.

### VIII. Gauge Invariance and the vacuum current.

The fact that there are problems with gauge invariance and quantum field theory has a long history. A direct calculation, using perturbation theory, of the vacuum current,

induced by an externally applied electric potential does not produce a gauge invariant

result. This has been shown by many authors (see Pauli and Villars[7], section 22 of

Heitler[5], chap. 14 of Greiner[4], and Sakurai[8]). The calculation in Section VII

showing that $\vec{I}(\vec{x}, \vec{y}) \neq 0$ is consistent with these previous results.

This has always been recognized as a problem because quantum mechanics must be gauge invariant. The main approach to dealing with this problem seems to be to assume there is something "deviant" about the mathematics involved in this calculation. The non-gauge invariant terms are then isolated and removed from the calculation. This may involve some form of regularization (Ref [7]) where other functions are introduced that happen to have the correct behavior so that the non-gauge invariant terms are cancelled. However, as has been pointed out by Pauli and Villars [7], there is no physical explanation for introducing these functions. They are mathematical devices used to force the desired (gauge invariant) result.

In this section it will be shown that for the vacuum current to be gauge invariant the Schwinger term $\vec{I}(\vec{x},\vec{y})$ must vanish. And, as was shown in the last section, this term does not vanish if $|0\rangle$ is used as the vacuum state.

From Eq. 8.3 of Pauli [3] the first order change in the vacuum current due to an external perturbing electric potential is given by

$$\vec{J}_{vac}^{(1)}(\vec{x},t) = i\left\langle 0\left|\left[\hat{\vec{J}}(\vec{x},t), \int d\vec{y} \int_{-\infty}^{t} dt'\left(-\hat{\vec{J}}(\vec{y},t')\cdot\vec{A}(\vec{y},t') + \hat{\rho}(\vec{y},t')A_o(\vec{y},t')\right)\right]\right|0\right\rangle \quad (109)$$

In the above expression the operators $\hat{\vec{J}}(\vec{x},t)$ and $\hat{\rho}(\vec{x},t)$ are the current and charge operators, respectively, in the interaction representation. They are related to the Schrodinger operators by

$$\hat{\vec{J}}(\vec{x},t) = e^{i\hat{H}_ot}\hat{\vec{J}}(\vec{x})e^{-i\hat{H}_ot} \text{ and } \hat{\rho}(\vec{x},t) = e^{i\hat{H}_ot}\hat{\rho}(\vec{x})e^{-i\hat{H}_ot} \quad (110)$$

According to Eq. 3.11 of Pauli [3] the above interaction operators satisfy

$$\frac{\partial\hat{\rho}(\vec{x},t)}{\partial t} = -\vec{\nabla}\cdot\hat{\vec{J}}(\vec{x},t) \quad (111)$$

The change in the vacuum current $\delta \vec{J}_{vac}^{(1)}(\vec{x},t)$ due to a gauge transformation is obtained by using Eq. (5) in (109) to yield

$$\delta \vec{J}_{vac}^{(1)}(\vec{x},t) = i \left\langle 0 \left| \left[ \hat{\vec{J}}(\vec{x},t), \int d\vec{y} \int_{-\infty}^{t} dt' \left( \hat{\vec{J}}(\vec{y},t') \cdot \vec{\nabla}\chi(\vec{y},t') + \hat{\rho}(\vec{y},t') \frac{\partial \chi(\vec{y},t')}{\partial t'} \right) \right] \right| 0 \right\rangle \quad (112)$$

If quantum field theory is gauge invariant then a gauge transformation of the electric potential should produce no change in any observable quantity. Therefore $\delta \vec{J}_{vac}^{(1)}(\vec{x},t)$ should be zero. To verify this we will solve the above equation as follows.

$$\int_{-\infty}^{t} dt' \hat{\rho}(\vec{y},t') \frac{\partial \chi(\vec{y},t')}{\partial t'} = \Big|_{-\infty}^{t} \hat{\rho}(\vec{y},t')\chi(\vec{y},t') - \int_{-\infty}^{t} dt' \chi(\vec{y},t') \frac{\partial \hat{\rho}(\vec{y},t')}{\partial t'} \quad (113)$$

Assume that $\chi(\vec{y},t)=0$ at $t \to -\infty$. Use this and Eq. (111) in the above expression to obtain

$$\int_{-\infty}^{t} dt' \hat{\rho}(\vec{y},t') \frac{\partial \chi(\vec{y},t')}{\partial t'} = \hat{\rho}(\vec{y},t)\chi(\vec{y},t) + \int_{-\infty}^{t} dt' \chi(\vec{y},t') \vec{\nabla} \cdot \vec{J}(\vec{y},t') \quad (114)$$

Substitute this into Eq. (112) to obtain

$$\delta \vec{J}_{vac}^{(1)}(\vec{x},t) = i \left\langle 0 \left| \left[ \hat{\vec{J}}(\vec{x},t), \int d\vec{y} \int_{-\infty}^{t} dt' \left( \hat{\vec{J}}(\vec{y},t') \cdot \vec{\nabla}\chi(\vec{y},t') + \chi(\vec{y},t') \vec{\nabla} \cdot \vec{J}(\vec{y},t') \right) \right] \right| 0 \right\rangle$$
$$+ i \left\langle 0 \left| \left[ \hat{\vec{J}}(\vec{x},t), \int \hat{\rho}(\vec{y},t)\chi(\vec{y},t)d\vec{y} \right] \right| 0 \right\rangle \quad (115)$$

Rearrange terms to obtain

$$\delta \vec{J}_{vac}^{(1)}(\vec{x},t) = i \left\langle 0 \left| \left[ \hat{\vec{J}}(\vec{x},t), \int_{-\infty}^{t} dt' \int d\vec{y} \vec{\nabla} \cdot \left( \hat{\vec{J}}(\vec{y},t')\chi(\vec{y},t') \right) \right] \right| 0 \right\rangle$$
$$+ i \left\langle 0 \left| \left[ \hat{\vec{J}}(\vec{x},t), \int \hat{\rho}(\vec{y},t)\chi(\vec{y},t)d\vec{y} \right] \right| 0 \right\rangle \quad (116)$$

Assume reasonable boundary conditions at $|\vec{y}| \to \infty$ so that

$$\int d\vec{y} \vec{\nabla} \cdot \left( \hat{\vec{J}}(\vec{y}, t') \chi(\vec{y}, t') \right) = 0 \tag{117}$$

Use this to obtain

$$\delta \vec{J}_{vac}^{(1)}(\vec{x}, t) = i \langle 0 \| \left[ \hat{\vec{J}}(\vec{x}, t), \int \hat{\rho}(\vec{y}, t) \chi(\vec{y}, t) d\vec{y} \right] \| 0 \rangle$$
$$= i \int \langle 0 \| \left[ \hat{\vec{J}}(\vec{x}, t), \hat{\rho}(\vec{y}, t) \right] \| 0 \rangle \chi(\vec{y}, t) d\vec{y} \tag{118}$$

Use Eq. (110) and the fact that $\hat{H}_o | 0 \rangle = 0$ in the above to obtain

$$\delta \vec{J}_{vac}^{(1)}(\vec{x}, t) = i \int \langle 0 \| \left[ \hat{\vec{J}}(\vec{x}), \hat{\rho}(\vec{y}) \right] \| 0 \rangle \chi(\vec{y}, t) d\vec{y} = -i \int \vec{I}(\vec{x}, \vec{y}) \chi(\vec{y}, t) d\vec{y} \tag{119}$$

Therefore for $\delta \vec{J}_{vac}^{(1)}(\vec{x}, t)$ to be zero, for arbitrary $\chi(\vec{y}, t)$, the Schwinger term $\vec{I}(\vec{x}, \vec{y})$ must be zero. When $| 0 \rangle$ is replaced by $| 0, \Delta E_w \to \infty \rangle$ in Eq. (109) then $\vec{I}(\vec{x}, \vec{y})$ is replaced by $\vec{I}'(\vec{x}, \vec{y})$ in Eq. (119). Since $\vec{I}'(\vec{x}, \vec{y})$ has been shown to be zero then the vacuum current will be gauge invariant if $| 0, \Delta E_w \to \infty \rangle$ is used as the vacuum state.

## IX. Conclusion

In this article we have shown that the state vector $| 0 \rangle$ is not consistent with the requirements for gauge invariance. One consequence of using $| 0 \rangle$ for the vacuum state is that the Schwinger term is not zero. This term must vanish for the theory to be gauge invariant. The result is that non-gauge invariant terms appear in the calculation of the vacuum current. This has led previous researchers to speculate that the reason for these terms was due to some difficult or deviant mathematical behavior in the calculation of the vacuum current. Here we have shown that the reason for these non-gauge invariant terms is not deviant mathematical behavior but simply that, as was shown in Section III, gauge invariance requires the possibility that quantum states can exist whose free field energy is less than that of the vacuum state. Based on this fact one would expect that a calculation of the vacuum current using $|0\rangle$ as the vacuum state would, indeed, produce a non-gauge invariant result. The non-gauge invariant terms are not a result of deviant or difficult mathematics but are the correct result of a straight forward mathematical calculation.

On the other hand, as has been shown, the state vector $|0, \Delta E_w \to \infty\rangle$ is consistent with the requirements of gauge invariance. It is similar to $|0\rangle$ in that is stabilizes positive energy particles against transitions into negative energy states and it is consistent with positron theory. However, it allows for the existence of negative energy states by including negative energy basis states which is necessary for a gauge invariant theory.

<div align="center">References</div>